\documentclass[twocolumn,floatfix,prl]{revtex4-1}
\pdfoutput=1
\usepackage[pdftex]{graphicx}

\newcommand{\APC}{APC, AstroParticule et Cosmologie, Universit\'{e} Paris Diderot, CNRS/IN2P3, CEA/IRFU, Observatoire de Paris, Sorbonne Paris Cit\'{e}, 75205 Paris Cedex 13, France}
\newcommand{\CEA}{Commissariat \`{a} l'Energie Atomique et aux Energies Alternatives, Centre de Saclay, IRFU, 91191 Gif-sur-Yvette, France}

\begin{document}
\title{Optimization for Mass Hierarchy}
\author{John LoSecco}
\affiliation{\APC}
\affiliation{\CEA}
\email{losecco@nd.edu}
\date{\today}
\begin{abstract}
The $\Delta m^{2}_{13}$ oscillation frequency for reactor neutrinos differs
by 6.4\% between normal
and inverted mass hierarchy.  This frequency difference accumulates to a phase
difference over distance and time.
The optimal distance is when the maximum phase difference between hierarchies
occurs near the peak in the observable reactor neutrino spectrum.
\end{abstract}
\keywords{neutrino oscillations, mass hierarchy, optimize}
\maketitle
Recent developments in neutrino mixing include measurement of
$\theta_{13}$~\cite{DC,DBRENO} and refinement of parameters~\cite{Minos}.
Outstanding questions
include the mass hierarchy, the CP violating phase $\delta$, sterile neutrinos,
\begin{figure}
\includegraphics[width=0.41\textwidth]{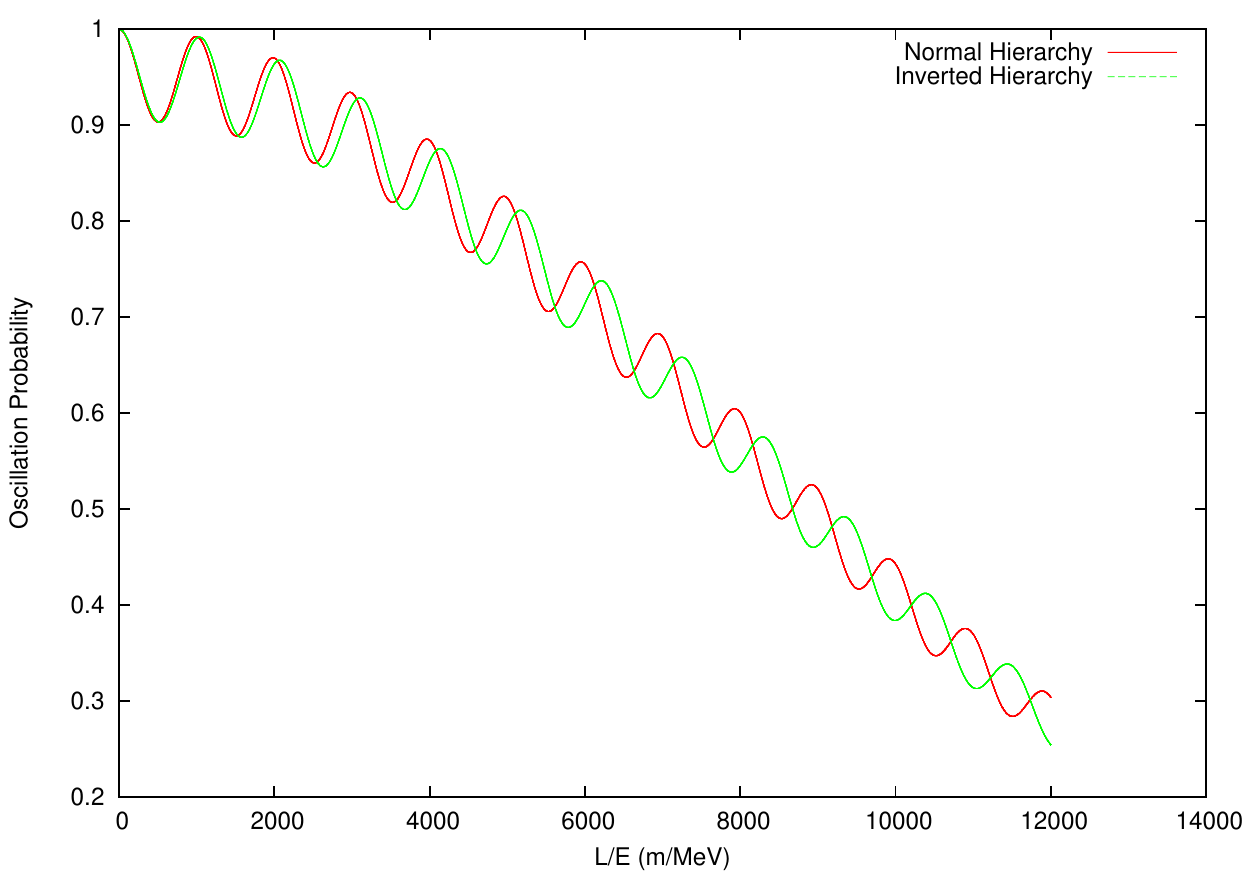}
\caption{\label{VacLE} The vacuum oscillation probability for neutrino
survival as a function of L/E for the normal and inverted mass hierarchies.
The frequency difference shows up as a phase difference after many
oscillations.}
\end{figure}
the Majorana nature of neutrino mass
and the overall neutrino mass scale.  Experiments are underway to better
understand these questions.
\begin{figure}
\includegraphics[width=0.41\textwidth]{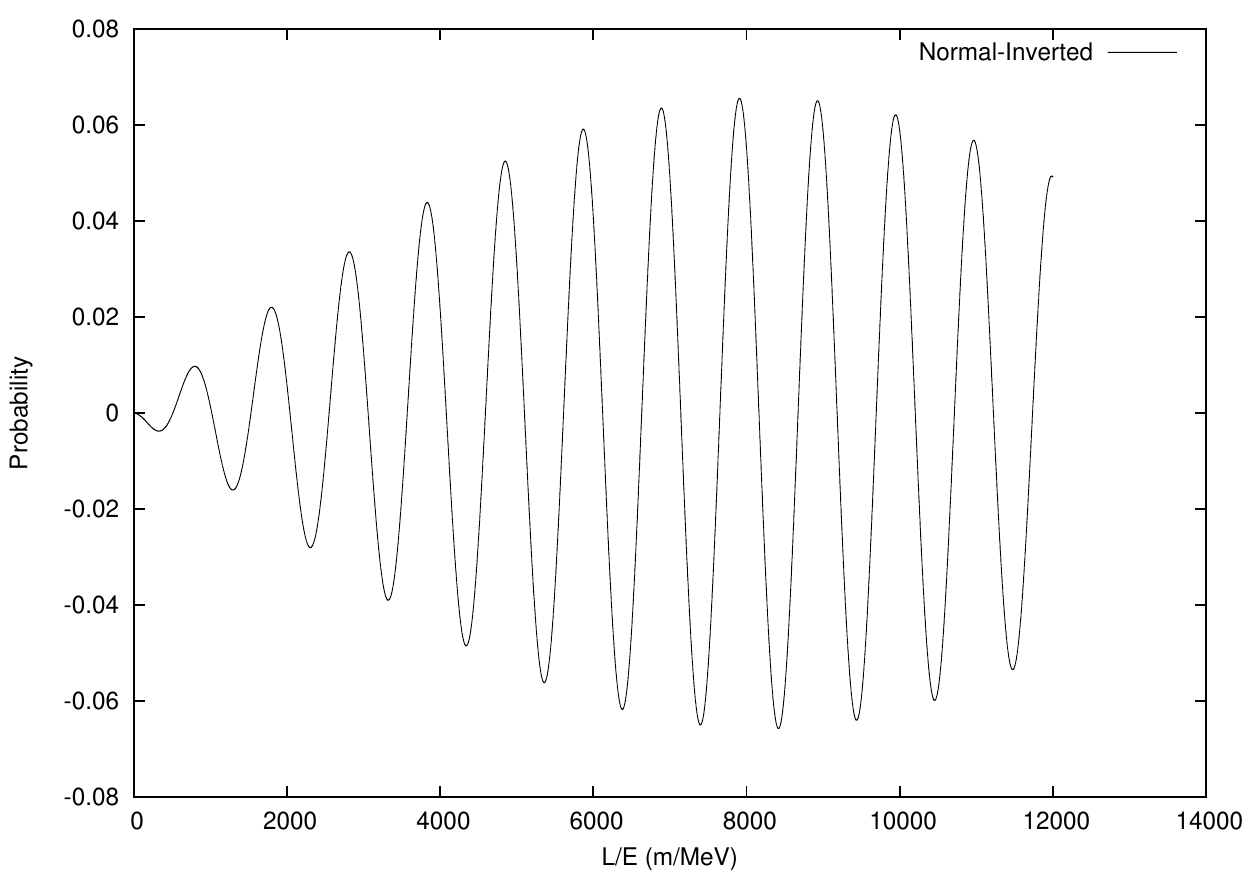}
\caption{\label{Norm-Inv} The transition probability  difference between
the normal and inverted mass hierarchy as a function of L/E in m/MeV.}
\end{figure}

The measurement of $\theta_{13}$ has done a great deal to permit the field
to expand quickly.  Plans to measure the mass hierarchy frequently involve
ambiguities with $\delta$.  Given the unexpectedly large value for
$\theta_{13}$ it is worth reconsidering strategies to determine the
mass hierarchy.

The transition probability to find a neutrino of type $\beta$ after a time $t$
when starting with a neutrino of type $\alpha$ in vacuum is given
by~\cite{PDG2012}:
\[
P_{\alpha \rightarrow \beta} = | <\nu_{\beta}|\nu_{\alpha}(t)>|^{2}=
| \Sigma_{j} U^{*}_{\alpha j} U_{\beta j} e^{-i m^{2}_{j}L/2E} |^{2}
\]
Where $U_{\beta j}$ are elements of the complex PMNS matrix, the $m^{2}_{j}$
are the square of the masses of the $j$'th neutrino mass eigenstate and
$L \approx c t$.
$\overline{\nu_{e}}$ disappearance experiments to measure the mass hierarchy have an
advantage that the measurement is independent of CP violating phases so
those ambiguities can be avoided.

In the case of an electron antineutrino disappearance experiment this can
be written as~\cite{Learned}:
\begin{eqnarray} \label{OscEqn}
P_{e \rightarrow e}=1-
( \cos^{4}(\theta_{13}) \sin^{2}(2 \theta_{12}) \sin^{2}(\Delta_{21})\\
+ \cos^{2}(\theta_{12}) \sin^{2}(2 \theta_{13}) \sin^{2}(\Delta_{31}) \nonumber \\
+ \sin^{2}(\theta_{12}) \sin^{2}(2 \theta_{13}) \sin^{2}(\Delta_{32}) ) \nonumber
\end{eqnarray}
where $\Delta_{ij} = 1.267 ( m^{2}_{i} - m^{2}_{j} ) \frac{L}{E}$.  Since
$\cos^{2}(\theta_{12}) \approx 0.7$ and $\sin^{2}(\theta_{12}) \approx 0.3$
the high frequency oscillation is dominated by $\approx \Delta_{31}$.
The L/E plot
shows (figure \ref{VacLE}) a low amplitude high frequency oscillation
at $\approx \Delta_{31}$ added to a high amplitude low frequency oscillation at
$\Delta_{21}$.

\begin{figure}[t]
\includegraphics[width=0.41\textwidth]{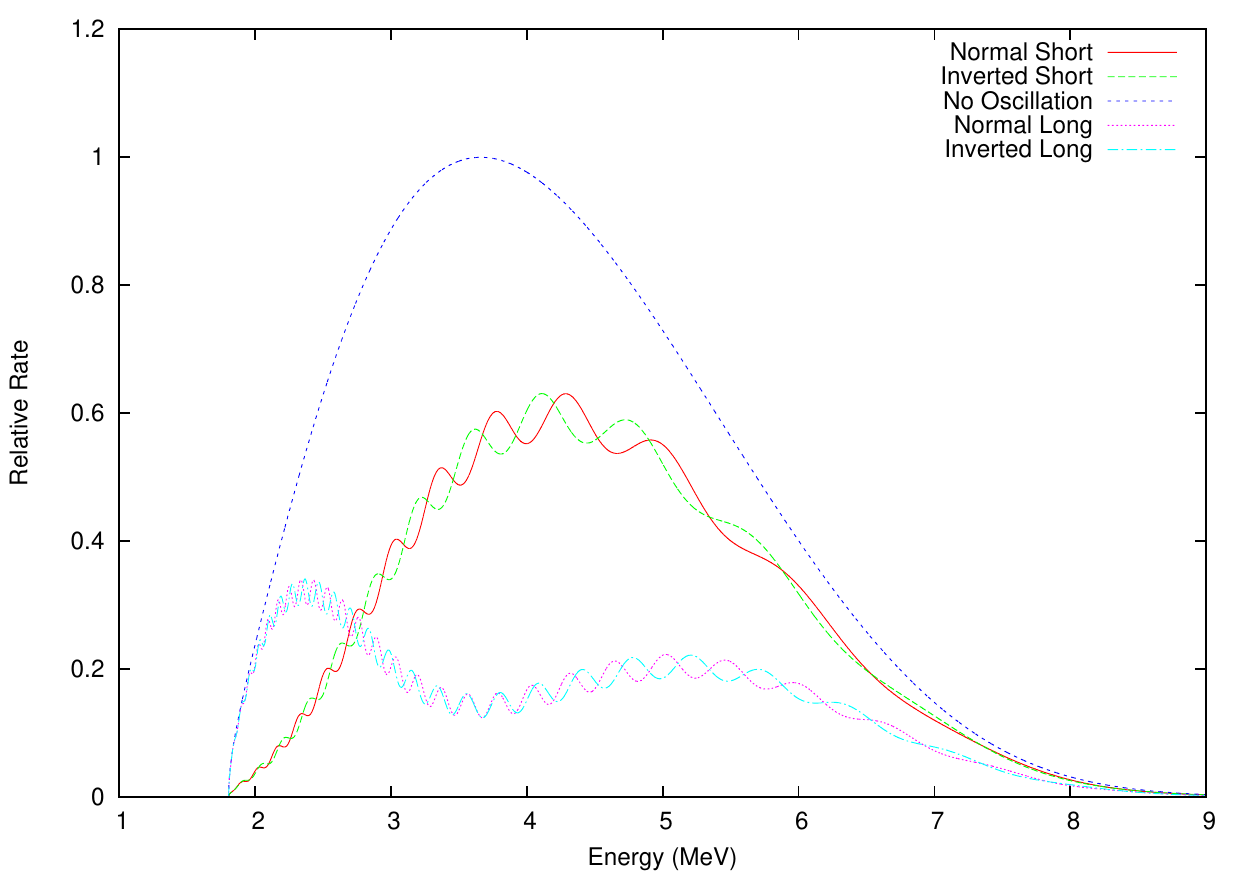}
\caption{\label{Spect} The modulated neutrino spectrum.  The unoscillated signal
is shown above in blue.  The oscillated signal for baselines of 60 km
and 30 km are shown for both mass hierarchies.}
\end{figure}

\begin{figure}[t]
\includegraphics[width=0.41\textwidth]{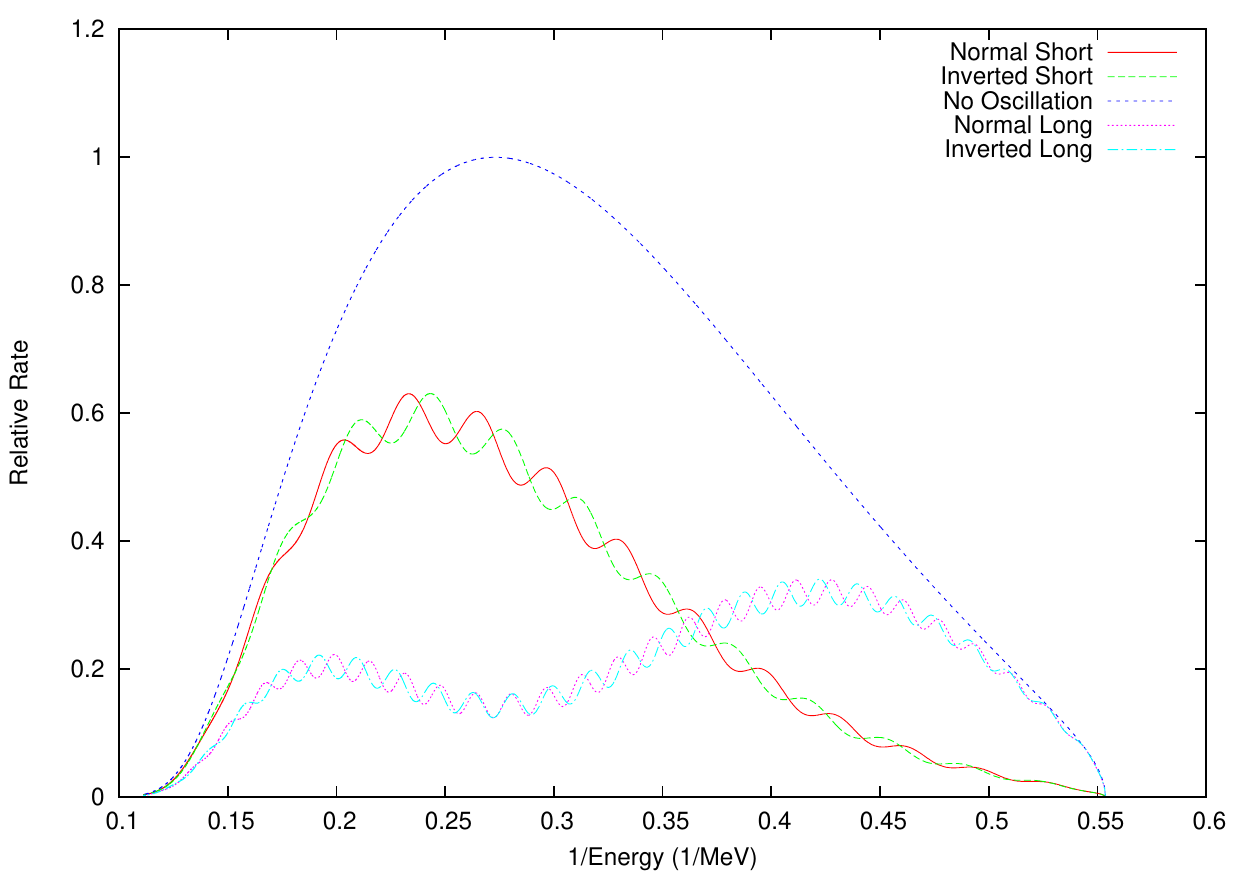}
\caption{\label{InvSpect} The modulated neutrino spectrum plotted as a function
of 1/E.  The curves are the same as those plotted in figure \ref{Spect}.
This illustrates the better energy separation between the peaks of the
$\theta_{13}$ oscillations at shorter baselines.  The even spacing in this plot
also makes it easier to see the relative phase of the normal and
inverted hierarchy oscillations.}
\end{figure}

A popular method to determine the mass hierarchy~\cite{Learned,Wang} is to
position a large reactor
antineutrino detector near the solar neutrino oscillation ($\Delta m^{2}_{12}$)
minimum and to
look at ripples in the spectrum caused by $\theta_{13}$ oscillations.
The combination of large distances and the oscillation minimum leads to
very low rates, resulting in the need for a very large detector and long
exposure times.  The oscillation frequency, the ripple spacing, also makes
serious demands on the detector resolution, on the order of 3\% to resolve
the ripples in the spectrum.

The portion of the transition probability sensitive to
the mass hierarchy can be isolated from equation \ref{OscEqn}.
\begin{eqnarray}
D=\sin^{2}(2 \theta_{13}) \cos^{2}(\theta_{12})  \sin^{2}(\Delta_{31}) \nonumber \\
=\frac{\sin^{2}(2 \theta_{13})}{2} \cos^{2}(\theta_{12}) (1- \cos(2 \Delta_{31})) \nonumber
\end{eqnarray}
Now $\Delta_{31} = \Delta_{32} + \Delta_{21}$ so
$\cos(2 \Delta_{31}) = \cos(2 \Delta_{32} + 2 \Delta_{21})
= \cos(2 \Delta_{32} ) \cos(2 \Delta_{21}) 
- \sin(2 \Delta_{32} ) \sin(2 \Delta_{21})$.
This gives:
\begin{eqnarray}
& & D=\frac{\sin^{2}(2 \theta_{13})}{2}\cos^{2}(\theta_{12}) \nonumber \\
& & (1-
(\cos(2 \Delta_{32} ) \cos(2 \Delta_{21}) 
- \sin(2 \Delta_{32} ) \sin(2 \Delta_{21}))) \nonumber
\end{eqnarray}

\begin{figure}[t]
\includegraphics[width=0.41\textwidth]{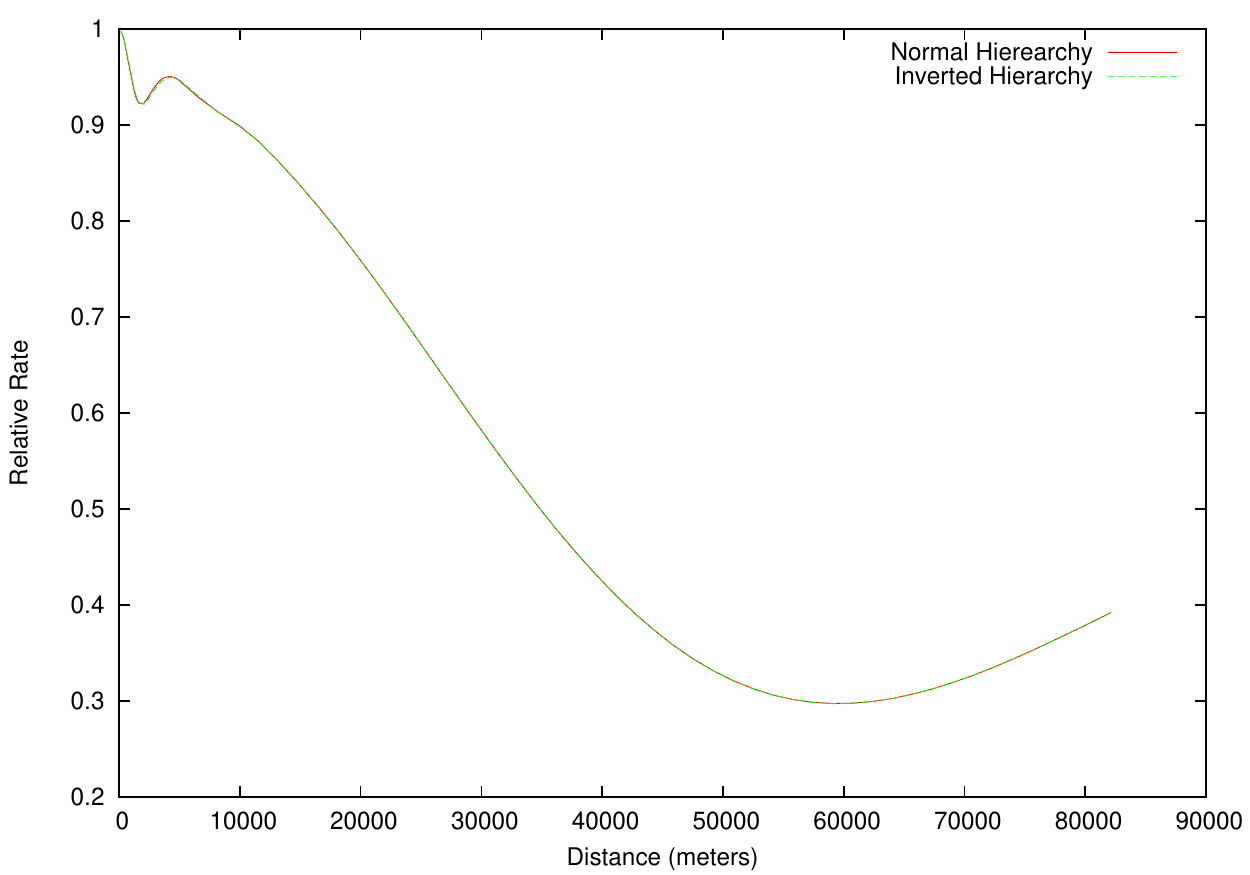}
\caption{\label{RatDist} The relative event rate due to oscillations
as a function of the distance to the reactor.  This is in addition to the
$\frac{1}{L^{2}}$ due to the drop in neutrino flux with distance.}
\end{figure}

The mass hierarchy is the sign of $\Delta_{32}$.  The only term in $D$ odd in
$\Delta_{32}$ is $\frac{\sin^{2}(2 \theta_{13})}{2} \cos^{2}(\theta_{12})
\sin(2 \Delta_{32} ) \sin(2 \Delta_{21})$.  The difference between normal
and inverted hierarchy (figure \ref{Norm-Inv}) is:
\[
|D_{N}-D_{I}|=\cos^{2}(\theta_{12})\sin^{2}(2 \theta_{13})
|\sin(2 |\Delta_{32}| ) \sin(2 \Delta_{21})|
\]
The maximum size of this difference is
$\cos^{2}(\theta_{12})\sin^{2}(2 \theta_{13})$.
Since the $\Delta_{21}$ oscillation is fairly slow this maximum difference can
be found near $|\sin(2 \Delta_{21} )|=1$.  The smallest value maximizing it is
$2 \Delta_{21} = \pi/2$.
For $\Delta m^{2}_{21}=7.54 \times 10^{-5}$ eV$^{2}$
this suggests an optimal L/E near 8200 m/MeV, figure \ref{Norm-Inv}.

\begin{eqnarray}
& & \sin(2 |\Delta_{32}| ) \sin(2 \Delta_{21}) = \nonumber \\
& & \frac{1}{2}
( \cos( 2 (|\Delta_{32}| - \Delta_{21})) - \cos( 2 (|\Delta_{32}| + \Delta_{21})))
\nonumber
\end{eqnarray}

The largest observable difference between the two mass hierarchies
occurs when the two predictions are 180 degrees out of phase.
\begin{eqnarray}
& 2 (|\Delta_{32}| - \Delta_{21}) = n \pi \nonumber \\
& 2 (|\Delta_{32}| + \Delta_{21}) = (n+1) \pi \nonumber \\
& 4 |\Delta_{32}| = (2n+1) \pi \nonumber \\
& 4 \Delta_{21} = \pi \nonumber
\end{eqnarray}
The two oscillation
frequencies for the two possible mass hierarchies differ by about 6.4\%
so the optimal phase difference would first occur at about 7.8 oscillations.
\[
\frac{|\Delta_{32}|}{\Delta_{21}}=2n+1
\]
Which gives $n$=15.6, L/E=8200 m/MeV.

The extrema of $|D_{N}-D_{I}|$ are the solutions to the equation
\[
\frac{\tan(2 \Delta m^{2}_{12} L/E )}{\Delta m^{2}_{12}}
= - \frac{\tan(2 |\Delta m^{2}_{32}| L/E )}{|\Delta m^{2}_{32}|}
\]
A numerical search (figure \ref{Norm-Inv}) gives the L/E to the first
global maximum at
L/E=8418.  The smallest L/E which is over 90\% of this maximum
separation is at L/E=5861. 

The flux times cross section for a typical~\cite{DC2nd} reactor neutrino
spectrum peaks at about 3.66 MeV.  A neutrino propagation length of about
30 km would provide optimal conditions in the vicinity of this peak.
The actual shape of the spectrum is fuel dependent and depends on reactor
burnup so precise optimization is not possible.  But the broad nature of the
peak means that operating near the peak should be sufficient.  This paper
uses the fuel mix and cross section of the Double Chooz publication~\cite{DC2nd}
as typical.

Figure \ref{Spect} illustrates the effect of oscillations on a reactor neutrino
spectrum at two possible distances from the neutrino source.  Both the normal
and inverted hierarchy are shown for each distance.  At the {\em optimal}
distance the oscillation peaks and valleys are near the opposite feature for the
other hierarchy.

In addition to the much higher event rate occurring away from the oscillation
minimum and closer to the source, a shorter distance relaxes constraints
on the needed energy resolution.  Ripples in the $\frac{1}{E}$ distribution
occur with a frequency of $1.267 \Delta m^2 L$ MeV, where L is the source
to detector distance.  Smaller L gives better separation between the
normal and inverted hierarchy peak positions.  This is illustrated in figure
\ref{InvSpect}.

A shorter baseline for reactor neutrino experiments resolving the
neutrino mass hierarchy problem provides higher event rates and better
energy separation than running at the solar oscillation ($\Delta m^{2}_{12}$)
minimum.  In addition to the higher neutrino flux coming from
$\frac{1}{L^{2}}$ the event rate is higher due to the smaller effect of the
solar ($\Delta m^{2}_{12}$) oscillations, as illustrated in figure \ref{RatDist}.
The increase in rate due to oscillations is about a factor of 2.

\begin{figure}[t]
\includegraphics[width=0.41\textwidth]{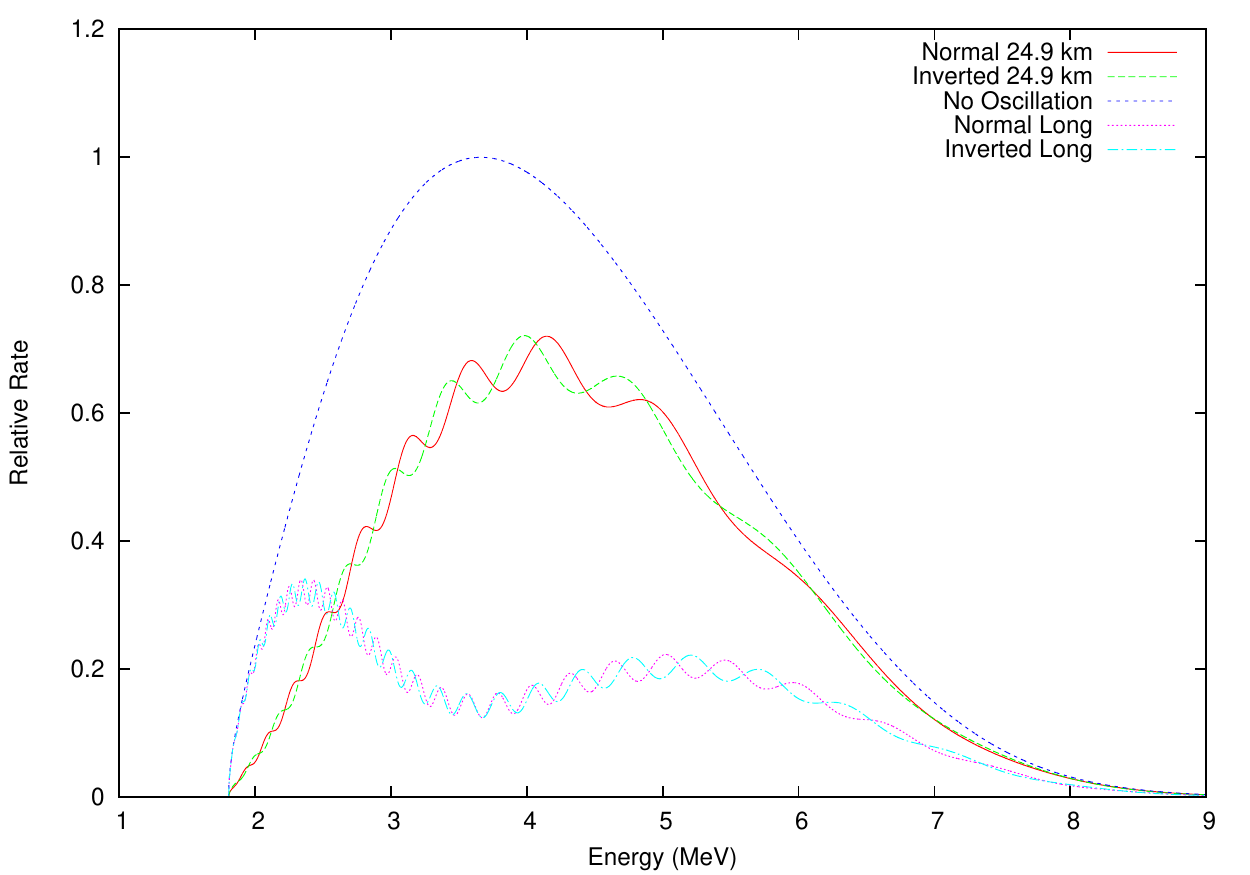}
\caption{\label{NotOpt}  The modulated neutrino spectrum.  The unoscillated
signal is shown above in blue.  The oscillated signal for baselines of 60 km
and 24.9 km are shown for both mass hierarchies.}
\end{figure}

The reactor neutrino flux times cross section is about 90\% of the peak value
at about 3.04 MeV.  A baseline of 24.9 km would give a 180 degree phase
difference between the mass hierarchies at this energy.  This is illustrated
in figure \ref{NotOpt}.  While not optimal such a baseline would still provide
good separation between the two possible hypotheses.
The oscillation enhancement is a factor of 2.3 at this location.
Optimizing at 3.66 MeV gives a factor of 2 in energy between the threshold and
the peak and a factor of 2.5 from the peak to the endpoint.  The broad nature
of the modulation (figure \ref{Norm-Inv}) indicates that most of the observable
spectrum would be sensitive to the mass hierarchy.

Since the neutrino mass parameters are only approximately known
the estimate given here is not precise.  But given the factor of 5 in the
accessible neutrino energy range the position optimization described here
should be adequate to get the optimal L/E very near the peak in the spectrum.
The value of $\Delta m^{2}$ in Fogli {\it et al.}~\cite{Minos}
has been used for our value of $\Delta m^{2}_{32}$.  Fogli {\it et al.}
has $\Delta m^{2}_{32} = \Delta m^{2} + \delta m^{2}/2$.   Most measurements
of $\Delta m^{2}_{32}$ come from muon neutrino disappearance
experiments~\cite{Minos2008}.

Systematic errors on $\Delta m^{2}_{32}$ may be problematic.  Since the
experiment can not measure the normal and inverted mass hierarchy and
compare them, comparison must be made to distributions based on an assumed
value of $\Delta m^{2}_{32}$ and a mass hierarchy.

I would like to thank Yifang Wang for useful discussions concerning the
Daya Bay II experiment.  I would like to thank Jamie Dawson and Didier Kryn
for useful comments on the manuscript.  This work was supported in part by
the Commission franco-am\'{e}ricaine.

After posting the first draft of this note I became aware of some recent work
on this question~\cite{Recent}.  I would like to thank J.~Evslin for
correspondence concerning the mass ambiguity.

\end{document}